\documentstyle[aps,prb
,twocolumn,epsf,bezier,floats
]{revtex}

\def\myabstract{
   \widetext\hsize
   \columnwidth\leftskip=0.10753
   \textwidth\rightskip\leftskip
   \nointerlineskip\small\relax
}

\begin{document}
\draft
\title{
   Free electron approximation for electron states of
   poly($p$-phenylene vinylene) and other conjugated systems.
}
\author{ D. L. Miller }
\address{
   Racah Institute of Physics,
   Hebrew University, Jerusalem, 91904,Israel\\
   {\rm\small Present address:}
  Dept. of Physics of Complex Systems,\\
 The Weitzmann Institute of science,
 Rehovot, 76100 Israel                \\
 e-mail  fndaniil@wicc.weizmann.ac.il  
}
\date{\today, file ary4pfeo.tex}
\twocolumn[
\maketitle
\begin{abstract}
\myabstract
   Free electron approximation for electron states in
   poly($p$-phenylene vinylene) and other conjugated systems
   is proposed. It provides simple and
   clear analytical expressions for energies of electron states and
   for wave functions. The results are in qualitative agreement with
   experiments. Our model does not contain fitting parameters.  We present
   two examples of the developed theory applications: exact calculation of
   electron energy in magnetic field, and scattering of electrons in
   copolymer.
\end{abstract}
\bigskip]

\narrowtext


\section{ Introduction }
\label{sec:intr}

The valence effective Hamiltonian method is used for many years as a
theoretical basis for band structure calculation of conjugated polymers, see
Ref.~\onlinecite{logdlund-93} and literature cited therein. However, very
simple approximate calculations can be performed in framework of introduced
by Schmidt\cite{schmidt-40} free electron model, see more references in book
of Salem\cite{salem66-fe}. This method is useful for classification of states
of one-dimensional chain molecules and aromatic molecules. In the latter case
it is known as Platt's Perimeter Model\cite{platt-1949} or Perimeter Free
Electron Orbital Theory(PFEO)\cite{pope82pfeo}. However, arylene based
polymers, like poly($p$-phenylene vinylene) (PPV) or poly(1,4-naphthalene
vinylene) (PNV), are not purely one-dimensional, because they contain
branching points. The generalization of free electron approach to this case
is the main purpose of the present work.
For the first time such calculations were undertaken by Ruedenberg and
Scherr\cite{Ruedenberg-Sherr-52} and here we develop the free electron network
model starting with symmetry arguments.

We also would like to consider the basic question why there exists
significant difference in characteristic energy scales of polyacetylene and
PPV $\pi$-electron bands structure. The answer was given in the work of Soos
{\em et al}\cite{soos-jul92}.  In the cited work authors addressed the small
energy scale of the band structure of PPV to the presence of phenyl
rings. The bands of PPV are formed from the levels of isolated phenyl ring
(benzene molecule).  The band width is determined by the coupling between
the rings, which can be obtained by the canonical transformation from the
coupling between adjacent carbon atoms.

In the present work we calculated the structure of PPV $\pi$-bands by means
of constructing of $\pi$-states from free electron wave functions. In our
model the proposed in Ref.\onlinecite{soos-jul92} topological gap  is the
consequence of the time-reversion symmetry of the wave function near the
triple nodes.  Triple node is carbon atom bound to three other carbons.
Therefore, the structure of $\pi$-bands of PPV is determined by the phases,
which are accumulated by the wave function having passed from one triple node
to another. Polymer deformations become unimportant in this model.   However,
the proposed here model is suitable only for organic structures build from
conjugated carbon chains.  Other kinds of organic structures cannot be
described by this model.

We obtained PPV $\pi$-electrons wave functions in simple analytical form and
their symmetry properties are explicitly seen. Symmetry of all PPV states is
essentially different from the symmetry of benzene molecule states, because
spatial oscillations period of free electron wave function changes
continuously with energy. One cannot say that this or that PPV electron state
originate from specific benzene molecule state. For example, almost all direct
light absorption processes between two highest occupied bands and two lowest
empty bands are dipole allowed, that agree with observed
experimentally\cite{halliday-93,cornil-jun94} four absorption bands.

Effect of magnetic field can be taken in to account exactly in framework
of our model. In the same way one can explain magnetic susceptibility
anisotropy of benzene molecule\cite{salem66-mp}. Magnetic field induces ring
current in benzene molecule and this effect gives main contribution in
magnetic susceptibility anisotropy. In PPV magnetic field also induces
currents in phenyl rings, and they give main contribution into PPV
diamagnetic susceptibility. It is different from the susceptibility of
equivalent number of isolated phenyl rings, because magnetic field induced
shifts of electron energies in PPV and benzene molecule are different.

In pure PPV one can distinguish occupied valence band and empty conductance
band. In order to obtain certain concentration of free carriers, holes in the
valence band and electrons in conductance band, one can dope polymer with
impurities or create structural defects. Our model can be used for
calculation of electron and holes transmission probabilities through
structural defect, if it is build from conjugated carbon chains. In this work
we considered naphthalene molecule inserted in PPV instead of phenyl ring in
order to see how strong it reflects carriers.

The accurate mathematical formulation of our model is done in next section.
The PPV and PNV bands structure calculated too. Developed method application
to the calculation of the diamagnetic susceptibility is given in
Sec.~\ref{sec:magn}. An example of electron scattering on monomer of a
different kind embedded into polymer, is considered in Sec.~\ref{sec:cop}. In
the same section we give the explicit form of PPV wave functions. The results
are summarized in Sec.~\ref{sec:summ}.


\section{ Electron states of PPV and PNV. }
\label{sec:wave}

Perimeter Free Electron Orbital Theory(PFEO)\cite{pope82pfeo} is a good tool
for description of benzene molecule $\pi$-electrons states. In this
approximation the local potential of the atoms is ignored, and it's assumed
that electrons can move freely around the molecule ring. The electron wave
function for such state is $\psi = \exp(ikr)$, where $r$ is coordinate along
the circular ring, and $k$ is the wave number. The periodic boundary condition
results in $kL = 2\pi n$, where $L$ is the perimeter of the ring that is
approximately 8.3\AA. So this theory gives sequence of benzene $\pi$-electron
levels
\begin{equation}
   {\cal{E}} = {(2\pi\hbar)^2\over mL^2}n^2, \;\; n = 0, 1, 2, \ldots
\label{eq:intr.1}
\end{equation}
Since we have one $\pi$-electron per each carbon atom, these six electrons
occupy two lowest energy levels( levels with $n\ge1$ are twice degenerate).

The big surprise is that the results of the above model are in good
agreement with experiment in spite of benzene molecule complex structure.
For example, carbons create periodic potential $V(r)$ along benzene ring
perimeter. The first order correction to the energy of $n$-th state due to the
presence of this potential is given by its matrix element
$\langle n| V |n \rangle$. This matrix element will be different from zero if
the period of the potential is equal to multiplied by integer number half the
wave length of $n$-th state wave function. The period of $V(r)$ is $L/3$ due
to the alternation of the $C$-$C$ bonds lengths, and the matrix element
$\langle n| V |n \rangle$ is not zero for the state with $n=3,6,\ldots$, but
other states don't feel this potential. Therefore, the application of
Eq.~(\ref{eq:intr.1}) is justified for low energy states $n\le2$.

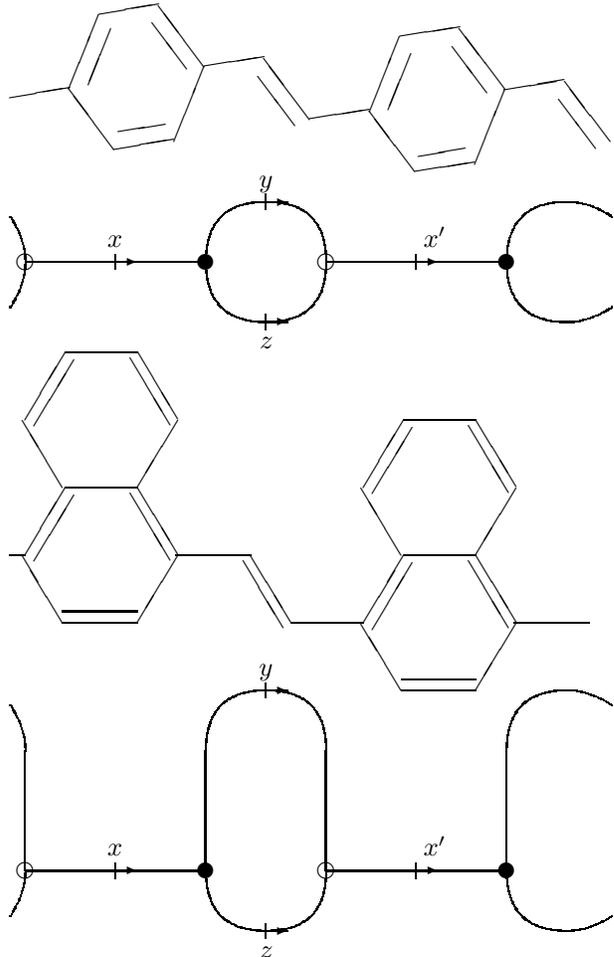
\begin{figure}
\unitlength=1.00mm
\linethickness{0.4pt}
\begin{picture}(81.00,127.36)
\put(3.00,93.00){\line(1,0){24.00}}
\put(43.00,93.00){\line(1,0){24.00}}
\bezier{64}(27.00,93.00)(27.00,101.00)(35.00,101.00)
\bezier{64}(35.00,101.00)(43.00,101.00)(43.00,93.00)
\bezier{64}(27.00,93.00)(27.00,85.00)(35.00,85.00)
\bezier{64}(35.00,85.00)(43.00,85.00)(43.00,93.00)
\bezier{64}(67.00,93.00)(67.00,101.00)(75.00,101.00)
\bezier{64}(67.00,93.00)(67.00,85.00)(75.00,85.00)
\bezier{28}(75.00,101.00)(78.00,101.00)(81.00,99.00)
\bezier{28}(75.00,85.00)(78.00,85.00)(81.00,87.00)
\bezier{28}(3.00,93.00)(3.00,96.00)(1.00,99.00)
\bezier{28}(1.00,87.00)(3.00,90.00)(3.00,93.00)
\put(27.00,93.00){\circle*{2.00}}
\put(67.00,93.00){\circle*{2.00}}
\put(43.00,93.00){\circle{2.00}}
\put(3.00,93.00){\circle{2.00}}
\put(56.00,93.00){\vector(1,0){2.00}}
\put(16.00,95.00){\makebox(0,0)[rb]{$x$}}
\put(36.00,103.00){\makebox(0,0)[rb]{$y$}}
\put(36.00,83.00){\makebox(0,0)[rt]{$z$}}
\put(56.00,95.00){\makebox(0,0)[lb]{$x'$}}
\put(35.00,102.00){\line(0,-1){2.00}}
\put(35.00,86.00){\line(0,-1){2.00}}
\put(35.00,101.00){\vector(1,0){2.99}}
\put(35.00,85.00){\vector(1,0){2.99}}
\put(55.00,94.00){\line(0,-1){2.00}}
\put(15.00,94.00){\line(0,-1){2.00}}
\put(15.00,93.00){\vector(1,0){3.01}}
\put(22.84,109.03){\line(2,5){3.98}}
\put(26.82,119.00){\line(-3,4){6.05}}
\put(8.84,116.03){\line(2,5){3.98}}
\put(14.82,108.00){\line(-3,4){6.05}}
\put(12.84,126.03){\line(6,1){8.00}}
\put(40.82,112.00){\line(-3,4){6.05}}
\put(26.84,119.03){\line(6,1){8.00}}
\put(40.84,112.03){\line(6,1){8.00}}
\put(14.84,108.03){\line(6,1){7.95}}
\put(62.84,106.03){\line(2,5){3.98}}
\put(66.82,116.00){\line(-3,4){6.05}}
\put(48.84,113.03){\line(2,5){3.98}}
\put(54.82,105.00){\line(-3,4){6.05}}
\put(52.84,123.03){\line(6,1){8.00}}
\put(54.84,105.03){\line(6,1){7.95}}
\put(1.00,114.88){\line(6,1){8.00}}
\put(66.68,115.91){\line(6,1){8.00}}
\put(19.92,124.87){\line(3,-4){4.63}}
\put(15.40,110.32){\line(6,1){6.27}}
\put(14.42,124.05){\line(-2,-5){3.00}}
\put(59.84,121.63){\line(3,-4){4.63}}
\put(55.31,107.08){\line(6,1){6.27}}
\put(54.33,120.82){\line(-2,-5){3.00}}
\put(33.97,117.73){\line(3,-4){4.93}}
\put(80.82,109.00){\line(-3,4){6.05}}
\put(73.97,114.73){\line(3,-4){4.93}}
\put(3.00,12.00){\line(1,0){24.00}}
\put(43.00,12.00){\line(1,0){24.00}}
\bezier{64}(27.00,28.00)(27.00,36.00)(35.00,36.00)
\bezier{64}(35.00,36.00)(43.00,36.00)(43.00,28.00)
\bezier{64}(27.00,12.00)(27.00,4.00)(35.00,4.00)
\bezier{64}(35.00,4.00)(43.00,4.00)(43.00,12.00)
\bezier{64}(67.00,28.00)(67.00,36.00)(75.00,36.00)
\bezier{64}(67.00,12.00)(67.00,4.00)(75.00,4.00)
\bezier{28}(75.00,36.00)(78.00,36.00)(81.00,34.00)
\bezier{28}(75.00,4.00)(78.00,4.00)(81.00,6.00)
\bezier{28}(3.00,28.00)(3.00,31.00)(1.00,34.00)
\bezier{28}(1.00,6.00)(3.00,9.00)(3.00,12.00)
\put(27.00,12.00){\circle*{2.00}}
\put(67.00,12.00){\circle*{2.00}}
\put(43.00,12.00){\circle{2.00}}
\put(3.00,12.00){\circle{2.00}}
\put(56.00,12.00){\vector(1,0){2.00}}
\put(16.00,14.00){\makebox(0,0)[rb]{$x$}}
\put(36.00,38.00){\makebox(0,0)[rb]{$y$}}
\put(36.00,2.00){\makebox(0,0)[rt]{$z$}}
\put(56.00,14.00){\makebox(0,0)[lb]{$x'$}}
\put(35.00,37.00){\line(0,-1){2.00}}
\put(35.00,5.00){\line(0,-1){2.00}}
\put(35.00,36.00){\vector(1,0){2.99}}
\put(35.00,4.00){\vector(1,0){2.99}}
\put(55.00,13.00){\line(0,-1){2.00}}
\put(15.00,13.00){\line(0,-1){2.00}}
\put(15.00,12.00){\vector(1,0){3.01}}
\put(3.00,54.00){\line(3,5){5.33}}
\put(8.33,63.00){\line(1,0){9.67}}
\put(18.00,63.00){\line(3,-5){5.33}}
\put(23.33,54.00){\line(-3,-5){5.33}}
\put(18.00,45.00){\line(-1,0){10.00}}
\put(8.00,45.00){\line(-3,5){5.33}}
\put(3.00,72.00){\line(3,5){5.33}}
\put(8.33,81.00){\line(1,0){9.67}}
\put(18.00,81.00){\line(3,-5){5.33}}
\put(23.33,72.00){\line(-3,-5){5.33}}
\put(8.00,63.00){\line(-3,5){5.33}}
\put(23.00,54.00){\line(1,0){10.00}}
\put(33.00,54.00){\line(3,-5){5.33}}
\put(38.33,45.00){\line(1,0){9.67}}
\put(48.00,45.00){\line(3,5){5.33}}
\put(53.33,54.00){\line(1,0){9.67}}
\put(63.00,54.00){\line(3,-5){5.33}}
\put(68.33,45.00){\line(-3,-5){5.33}}
\put(63.00,36.00){\line(-1,0){10.00}}
\put(53.00,36.00){\line(-3,5){5.33}}
\put(48.00,63.00){\line(3,5){5.33}}
\put(53.33,72.00){\line(1,0){9.67}}
\put(63.00,72.00){\line(3,-5){5.33}}
\put(68.33,63.00){\line(-3,-5){5.33}}
\put(53.00,54.00){\line(-3,5){5.33}}
\put(68.00,45.00){\line(1,0){10.00}}
\put(17.00,62.25){\line(3,-5){5.33}}
\put(17.00,80.25){\line(3,-5){5.33}}
\put(32.00,53.25){\line(3,-5){5.33}}
\put(62.00,53.25){\line(3,-5){5.33}}
\put(62.00,71.25){\line(3,-5){5.33}}
\put(4.21,53.16){\line(3,5){5.33}}
\put(4.21,71.16){\line(3,5){5.33}}
\put(49.21,44.16){\line(3,5){5.33}}
\put(49.21,62.16){\line(3,5){5.33}}
\put(18.00,46.48){\line(-1,0){10.00}}
\put(63.00,37.48){\line(-1,0){10.00}}
\put(3.00,12.00){\line(0,1){16.00}}
\put(27.00,12.00){\line(0,1){16.00}}
\put(43.00,12.00){\line(0,1){16.00}}
\put(67.00,12.00){\line(0,1){16.00}}
\put(1.00,54.00){\line(1,0){2.00}}
\end{picture}
   \caption{ Molecular structures of poly($p$-phenylene vinylene)
   and poly(1,4-naphthalene vinylene). Schematic diagrams show
   choice of coordinates along carbon chains for our model. The short
   vertical lines mark the origins of corresponding coordinates.}
\label{fig:chain}
\end{figure}

The main idea of this work is to extend the above approach to PPV, see
Fig.~\ref{fig:chain}, which can be considered as a sequence of benzene
molecules connected via short chain of two additional carbons. In the
framework of the considered model, electron runs along this short chain
unless it reaches the phenyl ring. Here electron has some probability to come
into the ring or to go back. The same picture repeats itself near the exit
from the phenyl ring. We will obtain the valid state of PPV if electron wave
function has the same amplitude after passing one monomer. Therefore,
each unit cell of the polymer is modeled by the three bonds, as it is shown
in Fig.~\ref{fig:chain}, where the length of each bond is $L/2$. The electron
can go from one bond to another at the nodes, which are marked by the
open and filled circles in Fig.~\ref{fig:chain}. The nodes correspond to
carbon atoms bound to three other carbons.

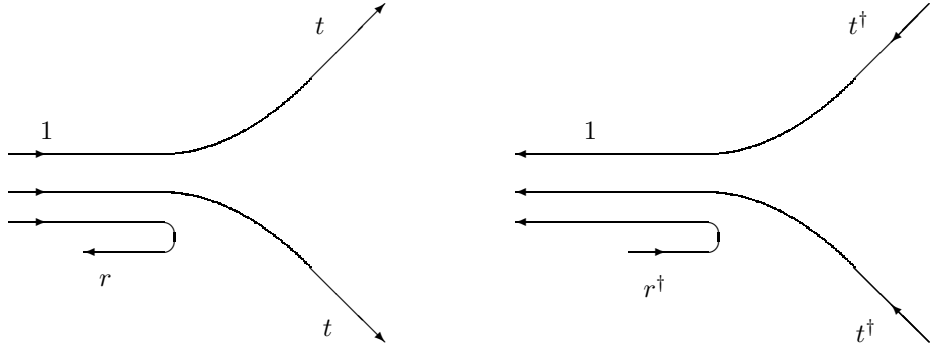
\begin{figure*}
\unitlength=1mm
\linethickness{0.4pt}
\centerline{
\begin{picture}(60.00,65.00)
\put(10.00,40.00){\line(1,0){20.00}}
\put(10.00,40.00){\vector(1,0){5.00}}
\put(50.00,30.00){\vector(1,-1){10.00}}
\bezier{96}(30.00,40.00)(40.00,40.00)(50.00,30.00)
\put(10.00,45.00){\line(1,0){20.00}}
\put(10.00,45.00){\vector(1,0){5.00}}
\put(50.00,55.00){\vector(1,1){10.00}}
\bezier{96}(30.00,45.00)(40.00,45.00)(50.00,55.00)
\put(10.00,36.00){\line(1,0){20.00}}
\put(10.00,36.00){\vector(1,0){5.00}}
\put(30.00,34.00){\oval(4.00,4.00)[r]}
\put(30.00,32.00){\vector(-1,0){10.00}}
\put(52.00,61.00){\makebox(0,0)[rb]{$t$}}
\put(53.00,23.00){\makebox(0,0)[rt]{$t$}}
\put(22.00,29.00){\makebox(0,0)[lt]{$r$}}
\put(15.00,47.00){\makebox(0,0)[cb]{$1$}}
\end{picture}
\hspace{1cm}
\begin{picture}(60.00,65.00)
\put(10.00,40.00){\line(1,0){20.00}}
\put(50.00,30.00){\line(1,-1){10.00}}
\put(50.00,30.00){\line(1,-1){10.00}}
\put(60.00,20.00){\vector(-1,1){5.00}}
\bezier{96}(30.00,40.00)(40.00,40.00)(50.00,30.00)
\put(10.00,45.00){\line(1,0){20.00}}
\put(50.00,55.00){\line(1,1){10.00}}
\put(60.00,65.00){\vector(-1,-1){5.00}}
\bezier{96}(30.00,45.00)(40.00,45.00)(50.00,55.00)
\put(10.00,36.00){\line(1,0){20.00}}
\put(30.00,32.00){\line(-1,0){10.00}}
\put(20.00,32.00){\vector(1,0){5.00}}
\put(30.00,34.00){\oval(4.00,4.00)[r]}
\put(52.00,61.00){\makebox(0,0)[rb]{$t^\dagger$}}
\put(53.00,23.00){\makebox(0,0)[rt]{$t^\dagger$}}
\put(22.00,29.00){\makebox(0,0)[lt]{$r^\dagger$}}
\put(15.00,47.00){\makebox(0,0)[cb]{$1$}}
\put(10.00,36.00){\vector(-1,0){5.00}}
\put(10.00,40.00){\vector(-1,0){5.00}}
\put(10.00,45.00){\vector(-1,0){5.00}}
\end{picture}
}
  \caption{ The wave coming to the node has the reflection
            amplitude $r$, and transmission amplitude to go to each
            other branch $t$, see left figure. 
            Inversion of time gives two conditions, see right figure. Amplitude of
            the wave going back has to be one and we obtain
            $1=tt^\dagger + tt^\dagger+rr^\dagger$. Amplitudes of the waves passing 
            to the branches has to be zero and
            therefore $0=tr^\dagger+tt^\dagger+rt^\dagger$. 
  }
\label{fig:ampl}
\end{figure*}

Let's imagine the free electron wave is propagating along the bond. When the
wave reaches the node the wave has amplitude $r$ to be reflected and
amplitude $t$ to go to each remaining bond, see Fig.~\ref{fig:ampl}. The amplitudes
to go left or right will be the
same if we assume that the node has $C_{3v}$ symmetry and the electron is in
the ground state of the bond confinement potential. More rigorously, let's
introduce coordinates $x$, $y$, $z$ along three bonds attached to one node,
black circle in Fig.~\ref{fig:chain}, with origins at the middle of the
corresponding bond. The electron wave functions in these bonds will be
\begin{eqnarray}
   Ae^{ikx} + Be^{-ikx} \;,
\nonumber\\
   Ce^{iky} + De^{-iky} \;,
\nonumber\\
   Ee^{ikz} + Fe^{-ikz} \;.
\label{eq:wave.2}
\end{eqnarray}
Each wave leaving the node has to be linear combination of all the three
waves coming in to this node,
\begin{eqnarray}
   Be^{-ikL/4} = rAe^{ikL/4} + tDe^{ikL/4} + tFe^{ikL/4} \;,
\nonumber\\
   Ce^{-ikL/4} = rDe^{ikL/4} + tFe^{ikL/4} + tAe^{ikL/4} \;,
\nonumber\\
   Ee^{-ikL/4} = rFe^{ikL/4} + tAe^{ikL/4} + tDe^{ikL/4} \;,
\label{eq:wave.3}
\end{eqnarray}
where phase $kL/4$ is acquired by electron having passed from the middle of
the bond to the node.

The amplitudes $t$ and $r$ are complex numbers defined by the properties of
the node. These amplitudes are not completely independent due to time
reversion symmetry, see Fig.~\ref{fig:ampl}. The simple calculation results in two 
conditions
\begin{eqnarray}
   |r-t|^2 = 1         \;,
\nonumber\\
   |r|^2 + 2|t|^2  = 1\;.
\label{eq:wave.8}
\end{eqnarray}
Therefore, two complex amplitudes $t$ and $r$ are defined by two real
parameters, which can be chosen as following
\begin{eqnarray}
   t &=& \sqrt{W\over2}e^{-i\theta/2}\;,
\nonumber\\
   r &=& -\biggl(\sqrt{W\over8}
         +i\sqrt{1-{9\over8}W}\biggr)e^{-i\theta/2}\;.
\label{eq:wave.9}
\end{eqnarray}
Here $0\le W \le 8/9$ and $0 \le \theta < 2\pi$. The time reversion symmetry
leads to some unusual property of ``triple'' node transmission and reflection
amplitudes. The absolute values of these amplitudes satisfy the conditions
\begin{equation}
  0 \le |t| \le 2/3\;,\;\; 1/3 \le |r| \le 1\;,
\label{eq:wave.9a}
\end{equation}
rather than the usual $0\le|r|\le1$, and $0\le|t|\le1$. Therefore, symmetric
``triple'' node always reflects electrons with the probability larger than
$1/9$, and this is consequence of the time reversion symmetry.

Our purpose is now to find the states in PPV. Since the polymer is periodic
structure, the electron wave function is characterized by quantum
number $\phi$. When electron pass one monomer, the wave function acquires
this additional phase. Let say that considered previously two bonds, with
coordinates $y$ and $z$ along them, join a new node, blank circle in
Fig.~\ref{fig:chain}, and make a loop. If the coordinate along third bond
attached to this node is $x'$ then the wave function in this bond will be
\begin{equation}
   e^{i\phi}\left\{Ae^{ikx'} + Be^{-ikx'} \right\}\;.
\label{eq:wave.4}
\end{equation}
Application of the transmission and reflection rules to this node gives us
three additional equations
\begin{eqnarray}
   Ae^{-ikL/4+i\phi} = rBe^{ikL/4+i\phi} + tCe^{ikL/4} + tEe^{ikL/4} \;,
\nonumber\\
   De^{-ikL/4} = rCe^{ikL/4} + tEe^{ikL/4} + tBe^{ikL/4+i\phi} \;,
\nonumber\\
   Fe^{-ikL/4} = rEe^{ikL/4} + tBe^{ikL/4+i\phi} + tCe^{ikL/4} \;.
\label{eq:wave.5}
\end{eqnarray}

Six equations, Eqs.~(\ref{eq:wave.3},\ref{eq:wave.5}), will be compatible
if the matrix determinant of their coefficients is zero. This determinant
has a following form, as a function of quantum number $\phi$ and energy,
which is taken into account by phase factor $\exp(ikL)$,
\begin{eqnarray}
   Q &=& \biggl[ e^{-ikL}-(r-t)^2 \biggr]
         \biggl[  (r+2t)^2(r-t)^2\biggr.
\nonumber\\
     &-& \Bigl(2r^2+2rt +t^2+4t^2\cos(\phi)\Bigr) e^{-ikL}
      + e^{-2ikL}
   \biggr]\;.
\nonumber\\
   \relax
\label{eq:wave.7}
\end{eqnarray}

\begin{mathletters}
The equation $Q=0$ gives the dispersion relation for electron states in the
polymer. It is a cubic equation for $\exp(ikL)$ and its three roots can be
represented in the following form
\begin{eqnarray}
   \cos(kL-\theta) &=& (9/4)W - 1\;,\;\; \sin(kL-\theta) > 0\;,
\nonumber \\ \relax \label{eq:wave.10} \\
   \cos(kL-\theta)  &=& W[5/4+\cos(\phi)] - 1\;,\;\;
   \sin(kL-\theta) > 0\;,
\nonumber \\ \relax \label{eq:wave.11}\\
   \cos(kL-\theta)  &=& W[5/4+\cos(\phi)] - 1\;,\;\;
   \sin(kL-\theta) < 0\;,
\nonumber \\ \relax \label{eq:wave.12}
\end{eqnarray}
where $k>0$ due to our choice of the reflection and transmission events.
The energies of the states are given by
\begin{eqnarray}
   {\cal{E}} = {\hbar^2k^2\over 2m}\;,
\label{eq:wave.14}
\end{eqnarray}
where $m$ is a free electron mass.
\label{eq:levels} \end{mathletters}

Equation~(\ref{eq:wave.10}) gives a sequence of isolated levels that
corresponds to waves trapped in rings. One can easily verify
that Eqs.~(\ref{eq:wave.3},\ref{eq:wave.5}) give for this case $A=B=0$, and
$C=-E$, $D=-F$.
Equations~(\ref{eq:wave.11},\ref{eq:wave.12}) give a sequence of bands
and envelop wave function is proportional to $e^{i\nu\phi}$, where $\nu$
is number of monomer in polymer chain. The amplitudes $A,B,C=E,D=F$ can be
obtained from Eqs.~(\ref{eq:wave.3},\ref{eq:wave.5}), where $k$ has to be
considered as a function of $\phi$ given by Eq.~(\ref{eq:wave.11}) or by
Eq.~(\ref{eq:wave.12}).

The comparison of the energy dispersions with numeric results of
Ref.\onlinecite{logdlund-93} shows that $\theta$ and $W$ are energy
independent and take the limit values: $\theta\approx0$, $W\approx8/9$. These
numbers mean that electron does not acquire additional phase after
passing triple node, and transmission of triple node is as good as it
is allowed by time reversion symmetry. Therefore, $\theta$ and $W$ take
their limit values and should not be considered as fitting parameters.
For these values\cite{graphite-bands} transmission and reflections amplitudes
are real and $|t|$ takes its maximal value, $t=2/3$, $r=-1/3$.

\begin{figure}
\unitlength=1mm
\begin{picture}(86,86)
\put(12.00,12.00){
   \put(0.0,0.0){\epsffile{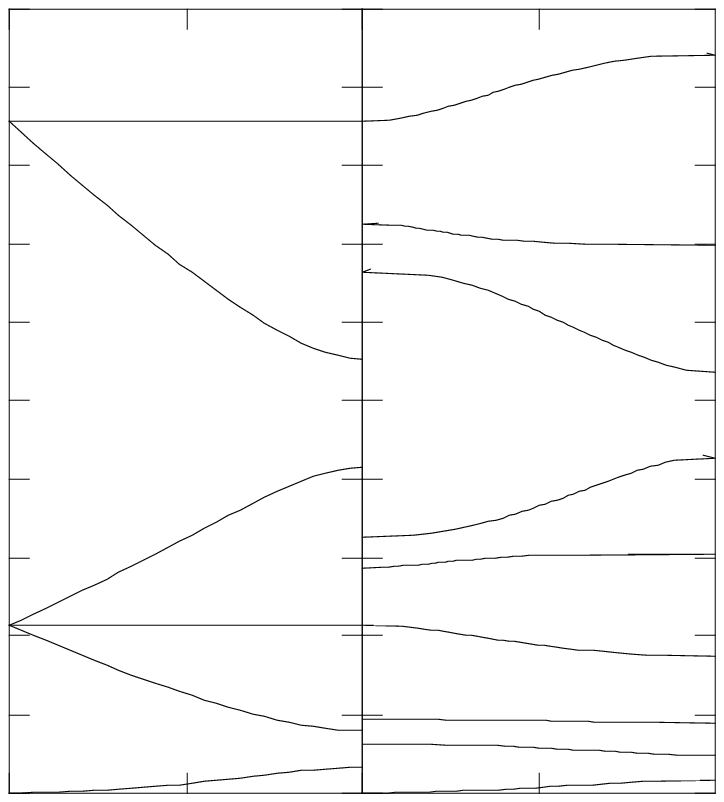}}
   \footnotesize
   \put(-1.0,41.00){\makebox(0,0)[rb]{$\cal E$,}}
   \put(-1.0,40.00){\makebox(0,0)[rt]{eV}}
   \put(-1.0,00.00){\makebox(0,0)[rc]{0}}
   \put(-1.0,08.00){\makebox(0,0)[rc]{1}}
   \put(-1.0,16.00){\makebox(0,0)[rc]{2}}
   \put(-1.0,24.00){\makebox(0,0)[rc]{3}}
   \put(-1.0,32.00){\makebox(0,0)[rc]{4}}
   \put(-1.0,48.00){\makebox(0,0)[rc]{6}}
   \put(-1.0,56.00){\makebox(0,0)[rc]{7}}
   \put(-1.0,64.00){\makebox(0,0)[rc]{8}}
   \put(-1.0,72.00){\makebox(0,0)[rc]{9}}
   \put(-1.0,80.00){\makebox(0,0)[rc]{10}}
   \put(18.00, -1.0){\makebox(0,0)[ct]{$\phi$}}
   \put(00.00, -1.0){\makebox(0,0)[lt]{0}}
   \put(35.00, -1.0){\makebox(0,0)[rt]{$\pi$}}
   \put(54.00, -1.0){\makebox(0,0)[ct]{$\phi$}}
   \put(37.00, -1.0){\makebox(0,0)[lt]{0}}
   \put(72.00, -1.0){\makebox(0,0)[rt]{$\pi$}}
   \put(10.00, 03.0){\makebox(0,0)[cc]{$\pi1$}}
   \put(10.00, 25.0){\makebox(0,0)[cc]{$\pi4$}}
   \put(30.00, 10.0){\makebox(0,0)[cc]{$\pi2$}}
   \put(30.00, 20.0){\makebox(0,0)[cc]{$\pi3$}}
   \put(30.00, 50.0){\makebox(0,0)[cc]{$\pi5$}}
   \put(30.00, 73.0){\makebox(0,0)[cc]{$\pi6$}}
   \put(45.00, 02.0){\makebox(0,0)[cc]{$\pi1$}}
   \put(45.00, 10.0){\makebox(0,0)[cc]{$\pi3$}}
   \put(45.00, 30.0){\makebox(0,0)[cc]{$\pi6$}}
   \put(45.00, 73.0){\makebox(0,0)[cc]{$\pi9$}}
   \put(65.00, 06.0){\makebox(0,0)[cc]{$\pi2$}}
   \put(65.00, 17.0){\makebox(0,0)[cc]{$\pi4$}}
   \put(65.00, 27.0){\makebox(0,0)[cc]{$\pi5$}}
   \put(65.00, 47.0){\makebox(0,0)[cc]{$\pi7$}}
   \put(65.00, 59.0){\makebox(0,0)[cc]{$\pi8$}}
}
\end{picture}
   \caption{The energies of calculated in the free electron approximation PPV
            and PNV
            $\pi$-bands. The only parameter in the calculations is perimeter
            of phenyl ring, $L=8.3$\AA.
            }
\label{fig:arybands}
\end{figure}

\begin{mathletters}
The energies of the isolated levels can be obtained from equation
$\cos(kL) = 1$, and the dispersed band energies can be found from equation
$\cos(kL) = [1+8\cos(\phi)]/9$.  The energies of six lowest bands, see
Fig.~\ref{fig:arybands}, are given by Eq.~(\ref{eq:wave.14}) and
\begin{eqnarray}
   \pi1: &\;\;&  kL = \arccos\left({1+8\cos(\phi)\over 9}\right)\;,
\label{eq:wave.15a} \\
   \pi2: &\;\;&  kL = 2\pi - \arccos\left({1+8\cos(\phi)\over 9}\right)\;,
\label{eq:wave.15b} \\
   \pi3: &\;\;&  kL = 2\pi\;,
\label{eq:wave.15c} \\
   \pi4: &\;\;&  kL = 2\pi + \arccos\left({1+8\cos(\phi)\over 9}\right)\;,
\label{eq:wave.15d} \\
   \pi5: &\;\;&
      kL = 4\pi - \arccos\left({1+8\cos(\phi)\over 9}\right)\;,
\label{eq:wave.15e} \\
   \pi6: &\;\;&  kL = 4\pi\;.
\label{eq:wave.15f}
\end{eqnarray}
The bands $\pi1$ -- $\pi4$ are occupied, and bands $\pi5$ and $\pi6$ are not.
Really, we have 8 $\pi$-electrons per unit cell and they occupy four lowest
bands. Fortunately, our model gives no states near $kL=3\pi/2$, where the
bands could be affected by the carbon chain periodic potential. However, the
calculated energy gap between the top of the valence, $\pi4$, band, where
$kL=3\pi-2\text{arcsin}(1/3)$, and the bottom of the conductance, $\pi5$,
band, where $kL=3\pi-2\text{arcsin}(1/3)$, is still smaller than the
experimentally measured value 2.4--2.5eV.
\label{eq:wave.15}
\end{mathletters}

The symmetry of all obtained above states is essentially different from the
benzene molecule states. It is because free electron wave vector $k$
determines the coordinates dependence of the PPV wave functions and it
changes continuously within the energy bands. The explicit form of the
calculated from Eqs.~(\ref{eq:wave.3},\ref{eq:wave.5}) PPV wave function
amplitudes  are given in Sec.~\ref{sec:wavefunctions}. One can
immediately see that direct optical transitions $\pi4\rightarrow\pi5$ and
$\pi3\rightarrow\pi6$ are dipole allowed\cite{soos-jul92} and their dipole
moment interband matrix element is parallel to PPV chain. Direct optical
transitions $\pi3\rightarrow\pi5$ and $\pi4\rightarrow\pi6$ are dipole
forbidden near the bands edges, $\phi=\pi$. The dipole moment interband
matrix element increases rapidly with transition energy, and therefore,
absorption bands corresponding to these processes also have to be
experimentally observed. Our theory predicts following density of states
peaks for four transitions: 1.4eV for $\pi4\rightarrow\pi5$, 3.4 for
$\pi3\rightarrow\pi5$, 4.4 for $\pi4\rightarrow\pi6$, and 6.4 for
$\pi3\rightarrow\pi6$. Three of these transtion energies are in qualitative agreement with
experimentally measured light absorption
spectrum\cite{halliday-93,cornil-jun94}, which have features near 2.5eV,
3.7eV, 4.8eV, and 6.5eV.

The similar picture is observed in PNV. It is more sophisticated conjugated
system synthesized recently.\cite{yoshino-sep93} The levels of naphthalene
molecule are also well described by the PFEO model\cite{pope82pfeo}.  In
other words $\pi$-electrons move freely around naphthalene molecule, see
Fig.~\ref{fig:chain}.  In this model naphthalene molecule is replaced by
two bonds, one of them having length $L/2$ and another having length $7L/6$.
Equations similar to Eqs.~(\ref{eq:wave.3},\ref{eq:wave.5}) can be written
for this model of PNV and from the condition of their compatibility we will
obtain the following band structure:
\begin{eqnarray}
    \cos(\phi) &=&\Bigl\{
    {9\over8}\cos(2kL) + {9\over8}\cos(5kL/3) + {9\over8}\cos(4kL/3)
\nonumber\\
    &+&
    {7\over8}\cos(kL) + {7\over8}\cos(2kL/3) - {1\over8}\cos(kL/3) \Bigr\}
\nonumber\\
    &/& \Bigl[
      \cos(kL)+\cos(2kL/3)+2\cos(kL/3)+1
    \Bigr] \;.
\nonumber\\
    \relax
\label{eq:cop.1}
\end{eqnarray}
This is algebraic equation of order five for $e^{ikL/3}$. Its solution leads
to sequence of bands, which are shown on Fig.~\ref{fig:arybands}. PNV has 12
carbon atoms in unit sell and six lowest bands are occupied. There are no
states near $kL=3\pi/2$ in this system, however the calculated energy gap
between valence, $\pi6$, and conductance, $\pi7$, bands is still smaller than
measured value\cite{yoshino-sep93} 2.3eV.  The conductance band offset in
possible PPV -- PNV system is much smaller than the gap, that is in
agreement with experiment.


\section{ Polymers in magnetic field. }
\label{sec:magn}

Let's consider experiment in which polymers are prepared in a two
dimensional layer. All the phenyl rings of the polymer will be
also lying in the same plane. If one applies magnetic field perpendicular to
that plane, electron propagating around phenyl ring will
acquire the additional phase, $\Phi$, which is equal to the
magnetic flux via the ring divided by $\hbar{c}/e$. This additional
phase can be easy introduced into Eqs.~(\ref{eq:wave.3},\ref{eq:wave.5}), and
the determinant of the coefficients matrix becomes
\begin{eqnarray}
   Q &=& \biggl\{ e^{-ikL}-(r-t)^2 \biggr\}
         \biggl\{  (r+2t)^2(r-t)^2\biggr.
\nonumber\\
     &-& \Bigl[
        2r^2+2rt - t^2 + 2t^2\cos(\Phi)
\nonumber\\
     &+& 4t^2\cos(\phi)\cos(\Phi/2)\Bigr]
     e^{-ikL} + e^{-2ikL} \biggr\}\;.
\nonumber\\
\relax
\label{eq:magn.1}
\end{eqnarray}
Again the equation $Q=0$ gives the dispersion relation for the
electron states in the polymer. The states $\pi3$ and $\pi6$ are not affected
by the magnetic field, this can be explained by the properties of these
states. In such a state electron is not only localized in the phenyl ring, it
is localized in one of the parts of the ring\cite{logdlund-93}. Therefore
electron in such a state does not go around the phenyl ring and does not
know about the magnetic field.

The electron energy in other $\pi$-states is periodic
function of magnetic flux
\begin{eqnarray}
   {\cal{E}}_{1,2}(\phi,\Phi) &=& {\hbar^2\over2mL^2}
   \Bigl[ \pi \mp \gamma(\phi,\Phi) \Bigr]^2\;,
   \\
\label{eq:magn.2a}
   {\cal{E}}_{4,5}(\phi,\Phi) &=& {\hbar^2\over2mL^2}
   \Bigl[ 3\pi \mp \gamma(\phi,\Phi) \Bigr]^2\;,
   \\
\label{eq:magn.2b}
   \cos\bigl[\gamma(\phi, \Phi)\bigr] &=&
       {1\over3}
       - {4\over9}\cos(\Phi)
\nonumber\\
       &-& {8\over9}\cos(\phi)\cos(\Phi/2)\;.
\label{eq:magn.3}
\end{eqnarray}
Here $\gamma(\phi, \Phi) > 0$. The energy of $\pi2$ and $\pi5$ states
decreases in weak magnetic field (paramagnetic states) and the energy of
$\pi1$ and $\pi4$ states increases (diamagnetic states).  For example, the
energy gap between the conduction and the valence bands decreases with
magnetic field, but the change is very small, since the area of the ring is
very small, see Ref.~\onlinecite{salem66-mp}, Eq.~4-37.

The above model does not take in to account the curvature of the polymers.
If the polymer has the points of the self crossing the electron states can
get additional magnetic moment that depends on the tunneling probabilities at
such points. The magnetic moment of $\pi$-states without considering of
this effect can be obtained from the derivatives of the energies with respect
to magnetic field.

The total magnetic moment of the polymer is given by the summation of the
magnetic moments of all the occupied states. The diamagnetic susceptibility
of the layered polymer material at zero temperature can obtained as a second
derivative of the total energy of the system with respect to magnetic field.
The approximate expression for the susceptibility per delocalized electron is
\begin{equation}
   \chi = N\int_{-\pi}^{\pi}{d\phi\over2\pi}
   \left( Se\over\hbar c\right)^2
   {\partial^2 \over\partial \Phi^2}
   \Bigl\{
      { {\cal{E}}_{2}(\phi,\Phi) + {\cal{E}}_{4}(\phi,\Phi)
      \over 2
      }
   \Bigr\}\;,
\label{eq:magn.5}
\end{equation}
where $S$ is phenyl ring area and $N$ is number of monomers. In this
expression we average susceptibility over delocalized states in the energy
interval corresponding to $\pi<kL<3\pi$. One can check that the averaged
susceptibility in interval $0<kL<\pi$ is the same. It is convenient to
express $\chi$ in terms of obtained from PFEO theory benzene molecule
susceptibility per electron\cite{pauling-1936}, $\chi_0=S^2e^2/(mc^2L^2)$.
Numeric integration with respect to $\phi$ gives $\chi \approx 0.7N \chi_0$.
Since PPV has six delocalized $\pi$ electrons per monomer, the PPV magnetic
susceptibility per mole of monomers has to be equal 0.7 of benzene molecule
molar diamagnetic susceptibility:
\begin{eqnarray}
   \chi_{\text{PPV}} = 0.7\chi_{\text{C$_6$H$_6$}}
\label{eq:magn.6}
\end{eqnarray}


\section{ Scattering of free electron waves in copolymer. }
\label{sec:wavefunctions}
\label{sec:cop}

The last example for the application of our almost heuristic theory to the
real physical problems is consideration of what will happen if we replace one
monomer in a polymer by monomer of a different kind. Let's consider
the copoly($p$-phenylene vinylene -- 1,4-naphthalene vinylene), which was used
in recent experiments\cite{davidov-95}.  This system can be obtained formally
from PPV if one replaces part of phenyl rings by naphthalene molecules. If
the naphthalene concentration is small, we can consider each monomer of
1,4-naphthalene vinylene as an impurity or a structural defect in PPV chain.
Therefore, such monomer partially reflects electrons propagating in PPV bands.

\begin{mathletters}
We start the calculation of naphthalene vinylene monomer transmission
coefficient with derivation of pure PPV wave functions. Three quantum numbers
$n$, $\kappa$, and $s\equiv\text{sign}(\gamma)$ specify each PPV $\pi$
electron state, see Eq.~(\ref{eq:stt.1}) below. Therefore, the wave functions
of all PPV $\pi$ states are given by the amplitudes $A=A_0(n,s,\kappa)$,
$B=B_0(n,s,\kappa)$, $C=C_0(n,s,\kappa)$, $D=D_0(n,s,\kappa)$,
$E=E_0(n,s,\kappa)$, $F=F_0(n,s,\kappa)$, which can be calculated for given
energy from Eqs.~(\ref{eq:wave.3},\ref{eq:wave.5}). Their explicit forms
are
\begin{eqnarray}
   A_0 &=& {1\over\sqrt{2L|\cosh(\xi)|}}
           e^{\xi/2}\;,
\label{eq:wave.18a}\\
   B_0 &=& {1\over\sqrt{2L|\cosh(\xi)|}}
           e^{-\xi/2}\;,
\label{eq:wave.18b}\\
   C_0 &=& E_0 = {i\over\sqrt{4L|\cosh(\xi)|}}
           e^{i\kappa/2+\xi/2}\;,
\label{eq:wave.18c}\\
   D_0 &=& F_0 = - {i\over\sqrt{4L|\cosh(\xi)|}}
           e^{i\kappa/2-\xi/2}\;,
\label{eq:wave.18d}\\
   \cosh(\xi) &=& 3\sin(\gamma/2)\;,
\label{eq:wave.21a}\\
   \sinh(\xi) &=& 2\sqrt{2}\sin(\kappa/2)\;,
\label{eq:wave.21b} \\
   \cos(\gamma/2) &=& {2\sqrt{2}\over3} \cos(\kappa/2)\;
\label{eq:wave.20} \\
   \phi &=&
   \left\{ \begin{array}{lll}
      \pi+\kappa,\;  &\phi>0,\; &\kappa<0, \\
      -\pi+\kappa,\; &\phi<0,\; &\kappa>0,
   \end{array} \right.
\\
   k_0L &=& (2n+1)\pi + \gamma \;,\;\;   -\pi\le\gamma\le\pi
\label{eq:stt.1i}
\end{eqnarray}
where the wave function is normalized on one particle per unit cell of
the polymer. In this system of equations free electron wave vector is also
the function of quantum numbers $k=k_0(n,s,\kappa)$. Vise versa, $k$
determines the quantities $\kappa,\gamma,\xi,n$, and thus all of them are
functions of energy.  Dimensionless wave number $\kappa$ is more convenient
than $\phi$, which was used in Eqs.~(\ref{eq:wave.15}) and
Fig.~\ref{fig:arybands}, because $\kappa$ goes to zero near the top of
valence band and near the bottom of conductance band.  There are two states
for each pair of quantum numbers $n$ and $\kappa$, one state with positive
$\gamma$, $s=1$, and one state with negative $\gamma$, $s=-1$.  For example
the states in the conductance and valence bands are given by $n=1$, with
$s=\pm1$ correspondingly. One can verify that the bands are parabolic near
the edges, where $|\kappa|\ll1$.
\label{eq:stt.1}
\end{mathletters}

Equation~\ref{eq:wave.20} is not necessary, but introduced for convenience.
When $\gamma$ is negative, the quantity $\xi$ has imaginary part, which is
equal to $i\pi$. Since dimensionless wave number $\kappa$ have to be real
Eq.~\ref{eq:wave.20} gives $\cos^2(\gamma/2)\le8/9$.  Therefore, the allowed
values of $\gamma$ lie in the intervals $-\pi\le\gamma\le-\gamma_0$ and
$\gamma_0\le\gamma\le\pi$, where $\gamma_0=2\arcsin(1/3)$. The state is
carrying electron, and flux is
\begin{equation}
   j = {\hbar k\over mL}\tanh(\xi)\;,
\label{eq:stt.4}
\end{equation}
which show the physical meaning of $\xi$.

Embedding of naphthalene vinylene monomer in to PPV chain result in
scattering of electrons and holes with dimensionless wave number $\kappa$. In
order to calculate transmission, $T$, and reflection, $R$, coefficients we
have to write equations similar to Eqs.~(\ref{eq:wave.3},\ref{eq:wave.5}) for
PNV free electron model, see Fig.~\ref{fig:chain}, and substitute
\begin{eqnarray}
   k  &=&  k_0(n,s,\kappa)\;,
\nonumber\\
   A  &=& A_0(n,s,\kappa) + RA_0(n,s,-\kappa) \;,
\nonumber\\
   A' &=& T B_0(n,s,\kappa)\;,
\nonumber\\
   B' &=& TB_0(n,s,\kappa)\;.
\nonumber\\
   B  &=& B_0(n,s,\kappa) + RB_0(n,s,-\kappa) \;,
\nonumber\\
   B' &=& TB_0(n,s,\kappa)\;.
\label{eq:cop.5}
\end{eqnarray}
Here one have to use $n=1$ and $s=\pm1$ for conduction and valence band
correspondingly. The further calculations are straightforward, the
qualitative results are the following. Near  both conduction and valence
bands edges transmission probability $|T|^2\approx0.8\kappa^2$, and therefore
it is proportional to the kinetic energy of quasiparticles. Far from the
edges it becomes of the order of one.


\section{ Summary. }
\label{sec:summ}

We have developed a new free electron model of PPV, which allows to calculate
band structure of polymer and hold wave functions of $\pi$ electrons in
simple analytical form.  We showed that the band structure contains only one
parameter, which is the averaged distance between two neighboring carbons.
The symmetry of PPV wave function allows direct optical transitions from
$\pi3$ and $\pi4$ states to $\pi5$ and $\pi6$ states, that explain four bands
observed on absorption spectrum. The calculated energy gap between $\pi4$ and
$\pi5$ states is not in agreement with experiment, but energies of three
other transitions are in agreement. This supports importance of many-body
polaronic\cite{furukawa95} and excitonic\cite{chandross-nov94,garstein-jul95}
effects for transitions near the energy gap. Similar models can be 
developed for other conjugated systems, for graphite planes or more 
complicated polymers.

We have calculated exactly the magnetic field dependence of electron 
energies, which is due to presence of phenyl rings. The diamagnetic 
susceptibility per monomer of PPV have been found approximately to be 0.7 of 
the susceptibility anisotropy of isolated benzene molecule. We also 
considered scattering of electrons on polymer structural defects.  We found 
that PPV electrons and holes transmission probability through
1,4-naphthalene vinylene monomer inserted in polymer is proportional to 
their kinetic energy. This is according to general property of scattering on 
potential occupiing final volume of space. We belive that proposed model
have many potential applications.

\acknowledgments

We wish to acknowledge prof. Dan~Davydov whose talk about conjugated polymers
initiated this work, prof. Boris~Laikhtman for discussion of this work and
criticism of the manuscript, and Slava Rotkin for discussion of the results.


\begin{references}


\bibitem{logdlund-93}
M.~L\"{o}gdlund, W.~R.~Salaneck, F.~Meyers, J.~L.~Br\'{e}das, G.~A.~Arbruckle,
  R.~H.~Friend, A.~B.~Holmes, G.~Froyer, Macromolecules {\bf 26},  3815
  (1993).

\bibitem{schmidt-40}
O.~Shmidt, Ber. Deut. Chem. Ges. {\bf 73A},  97  (1940).

\bibitem{salem66-fe}
L.~Salem, {\em The molecular orbital theory of conjugated systems.}
  (W.~A.~Benjamin, INC., New York, 1966), Chap.~7. Molecular orbital theory and
  the excited states of conjugated systems. UV Spectra.

\bibitem{platt-1949}
J.~R.~Platt, J.~Chem.~Phys. {\bf 17},  484  (1949).

\bibitem{pope82pfeo}
M.~Pope and C.~E.~Swenberg, {\em Electronic properties in organic crystals}
  (Oxford University Press, New York, 1982), Chap.~I.B. Molecular excited
  states., p.\ 7.

\bibitem{Ruedenberg-Sherr-52}
K.~Ruedenberg and C.~W.~Scherr, J. Chem. Phys. {\bf 21}, 1565(1953)

\bibitem{soos-jul92}
Z.~G.~Soos, S.~Etemad, D.~S.~Galv\~ao and S.~Ramasesha, Chem. Phys. Lett. {\bf
  194},  341  (1992).

\bibitem{halliday-93}
D.~A.~Halliday, P.~L.~Burn, R.~H.~Friend, D.~D.~C.~Bradley, A.~B.~Holmes, and
  A.~Kraft, Synth. Met. {\bf 55-57},  954  (1993).

\bibitem{cornil-jun94}
J.~Cornil, D.~Beljonne, R.~H.~Friend, J.~L.~Br\'edas, Chem. Phys. Lett. {\bf
  223},  82  (1994).

\bibitem{salem66-mp}
L.~Salem, {\em The molecular orbital theory of conjugated systems.}
  (W.~A.~Benjamin, INC., New York, 1966), Chap.~4. Molecular orbital theory of
  the Magnetic Properties of closed Shells.

\bibitem{graphite-bands}
We also made a calculation of a dispersion law of $\pi$-electrons of graphite
  planes and found from the fitting of our result to other data
  $\theta\approx0$ and $W\approx7.75/9$.

\bibitem{yoshino-sep93}
M.~Onoda, M.~Uchida, Y.~Ohmori, K.~Yoshino, Jpn. J. Appl. Phys. {\bf 23},  3895
   (1993).

\bibitem{pauling-1936}
L.~Pauling, J.~Chem.~Phys. {\bf 4},  673  (1936).

\bibitem{davidov-95}
H.~Hong, D.~Davidov, E.~Farage, H.~Chayet, R.~Neumann, and Y.~Avny,   , privet
  communication.

\bibitem{furukawa95}
Y.~Furukawa, Synth. Met. {\bf 69},  629  (1995).

\bibitem{chandross-nov94}
M.~Chandross, S.~Mazumdar, S.~Jeglinski, X.~Wey, Z.~V.~Vardeny, E.~W.~Kwock,
  and T.~M.~Miller, Phys. Rev. B {\bf 50},  14702  (1994).

\bibitem{garstein-jul95}
Yu.~N.~Garstein, M.~J.~Rice, and E.~M.~Conwell, Phys. Rev. B {\bf 52},  1683
  (1995).

\end{references}
\end{document}